# The breakdown of both strange metal and superconducting states at a pressure-induced quantum critical point in iron-pnictide superconductors

Shu Cai[1], Jinyu Zhao[1,4], Ni Ni[2,3], Jing Guo[1,5], Run Yang[1], Pengyu Wang[1,4], Jinyu Han[1,4], Sijin Long[1,4], Yazhou Zhou[1], Qi Wu[1], Xianggang Qiu[1,4,5], Tao Xiang[1,4], Robert J Cava[2], and Liling Sun[1,4,5]†

[1]*Institute of Physics, Chinese Academy of Sciences, Beijing 100190, China*
[2]*Department of Chemistry, Princeton University, Princeton, New Jersey 08544, USA*
[3]*Department of Physics and Astronomy, UCLA, Los Angeles, CA 90095, USA*
[4]*University of Chinese Academy of Sciences, Beijing 100190, China*
[5]*Songshan Lake Materials Laboratory, Dongguan, Guangdong 523808, China*

The strange metal (SM) state, characterized by a linear-in-temperature resistivity, is often seen in the normal state of high temperature superconductors. It is believed that the SM state is one of the keys to understand the underlying mechanism of high-$T_c$ superconductivity. Here we report the first observation of the concurrent breakdown of the SM normal state and superconductivity at a pressure-induced quantum critical point in an iron-pnictide superconductor, $Ca_{10}(Pt_4As_8)((Fe_{0.97}Pt_{0.03})_2As_2)_5$. We find that, upon suppressing the superconducting state by applying pressure, the power exponent ($\alpha$) changes from 1 to 2, and the corresponding coefficient $A^{\square}$, the slope of the temperature-linear resistivity per FeAs layer, gradually diminishes. At a critical pressure (~12.5 GPa), $A^{\square}$ and $T_c$ go to zero concurrently, where a quantum phase transition (QPT) from a superconducting state with a SM normal state to a non-superconducting Fermi liquid state takes place. Scaling analysis on the results obtained from the pressurized 1048 superconductor reveals that the change of $A^{\square}$ with $T_c$ obey the relation of $T_c \sim (A^{\square})^{0.5}$, which exhibits a similarity with that is seen in other chemically-doped unconventional superconductors, regardless of the type of the tuning method (doping or pressurizing), the crystal structure, the bulk or film superconductors and the nature of dopant (electrons or holes). These results suggest that there is a simple but powerful organizational principle of connecting the SM normal state with the high-$T_c$ superconductivity.

The strange metal (SM) state is an extraordinary normal state of high temperature superconductors, in which the electrical resistivity grows linearly with temperature in the low temperature limit. Such a unique state has been found in the doped cuprates, iron-pnictide superconductors[1-10], organic materials[8], heavy Fermion metals[11-13], infinite-layer nickelates[14] and twisted bilayer graphene[15]. Although many efforts have been made in past decades, the correlation of the SM state with the superconductivity still remains a subject of debate, which requires more experimental information from different experimental probes and materials. The central questions for this key issue are what is the intrinsic correlation between the SM normal state and the superconductivity, and what is the determining factor associated with the SM normal state for stabilizing high-$T_c$ superconductivity. To find clues for solving these puzzles, a superconductor with a pure SM normal state featured by $\alpha=1$ (in the form of $\rho = \rho_0 + AT^\alpha$) is needed. And better, if a superconducting system only hosts as few as possible competing orders in the tuning parameter - temperature phase diagram, then direct information to reveal the relation between these two may be obtained.

The iron-pnictide superconductor $Ca_{10}(Pt_4As_8)((Fe_{0.97}Pt_{0.03})_2As_2)_5$ is such an ideal superconducting material for this kind of study. The ground state of this superconductor is well understood, without the complication from ordered states, such as pseudogap state or a nearby long-range AFM state etc[16]. More intriguingly, it shows a $T$-linear resistivity behavior in its normal state at ambient pressure[16], providing us a unique platform for studying the correlation between the $T$-linear resistivity and the superconductivity.

Pressure is one of the non-thermal control parameters for tuning superconductivity because this method can shorten the interatomic distances and correspondingly leads to the change of the crystal and electronic structures without altering the chemistry. Moreover, it can avoid system's uncertainties associated with the differences between the samples with different doping levels, which poses significant limitations on the studies of chemically-doped single crystals. Thus, high pressure, as a 'clean' way, has been widely adopted as an independent control parameter to explore new phenomena and the evolution from one state to another in correlated electron systems[17-25]. In this study, we take the $Ca_{10}(Pt_4As_8)((Fe_{0.97}Pt_{0.03})_2As_2)_5$ superconductor (here referred to as the "1048 superconductor") as a target material to study the pressure-induced co-evolution of the SM and superconducting states.

The $Ca_{10}(Pt_4As_8)(Fe_2As_2)_5$ superconductor crystalizes in a tetragonal unit cell with -Ca-($Pt_4As_8$)-Ca-($Fe_2As_2$)-stacking[16]. Its structure can be descried as a distorted $CaFe_2As_2$ structure with every the other $Fe_2As_2$ layer replaced by a $Pt_4As_8$ layer, as shown in Fig.1a. At ambient pressure, its resistivity versus temperature shows a linear behavior above the superconducting transition at 26 K (Fig.1b), which closely resembles what is seen in optimally-doped cuprate superconductors[2,3,26,27]. When pressure is applied on the samples (S#1 and S#2) surrounded by the pressure transmitting medium (PTM) of NaCl, we find that $T_c$ shifts to low temperature upon compression and is invisible at 13 GPa and above for the sample #1 (Fig.1c and 1d), indicating that the application of pressure can suppress the superconductivity effectively. We repeat the measurements with new samples cut from different batches and obtain reproducible results (Fig.1e and 1f) - increasing pressure brings a monotonic reduction in $T_c$ and the superconductivity vanishes for the S#2 at 11.8 GPa. These results demonstrate the reproducible evolution of the superconductivity of the 1048 superconductor with pressure.

To clarify the possibility that the suppression of $T_c$ is related to the pressure inhomogeneity introduced by the solid pressure medium NaCl, we perform the high-pressure measurements on the S#4 with the liquid PTM of glycerin, and find the same results (Fig.1g and 1h) - $T_c$ declines with increasing pressure and vanishes at 12.8 GPa.

Then, we plot the resistance versus temperature for the pressurized S#1 and the S#4 in Fig.2, and make the actual fits to the temperature ($T$) dependence of the normal state resistance at low temperatures for the data obtained from the compressed samples based on the following equations:

$$R(T) = R_0' + A'T + B'T^2$$

$$R(T) = R_0 + AT^{\alpha}$$

Where $R_0$ and $R_0'$ are the residual resistance, $A'$, $B'$, $A$ and $\alpha$ are the slope of the $T$-linear resistance, the coefficient of $T$-square resistance, coefficient of power function and the power exponent of temperature, respectively (see Supplementary Information - SI).

Since the determination of the resistivity as a function of temperature at each pressure investigated requires the pressure dependence of sample thickness, which is difficult to be obtained in the high-pressure resistance measurements, we replace resistivity with resistance in the data fitting so as to take no account of the relation between the applied pressure and the sample thickness (see SI). It has been known that the high temperature superconductors such as cuprates and iron pnictides have a layered crystal structure, their superconductivities display a prominent two-dimensional characteristics[28-31]. To describe the correlation between the SM state and superconductivity reasonably, we employ $A^{\square}$, the $T$-linear resistivity coefficient normalized by the average distance between FeAs layers, to investigate this important correlation (see SI). The fitting is in good consistence with the experiment results (see red lines in Fig.2), giving a set of coefficient $A^{\square}$ that varies with pressure from 10.8 at 0.6 GPa to 0 at ~13 GPa and above for the S#1 (Fig.2a-2h), and from 12.8 at 1.1 GPa to 0 at~ 12.8 GPa and above for the S#4 (Fig. 2i-2p). The other fitting results can be found in SI. We extract the power exponent ($\alpha$), based on the equation in the form of power law, as a function of pressure and find that pressure changes $\alpha$ from 1 at ambient pressure to 2 at 13 GPa for the S#1 and 12.8 GPa for the S#4. When pressure is higher than ~13 GPa for the S#1 and 12.8 GPa for the S#4, the low temperature resistance shows curvature (Fig.2e-2h and 2m-2p) and approaches a $T^2$ dependence in the higher pressure range, where the superconductivity is fully suppressed. These results indicate that the application of pressure gradually drives the 1048 superconductor from a superconducting ground state with a SM normal state to a non-superconducting Fermi-liquid (FL) state.

We summarize the experimental results in the pressure-temperature phase diagram in Fig.3a. There are three distinct regions in the diagram: the superconducting (SC) state, the strange metal (SM) state (or non-Fermi liquid (NFL) behavior) and the non-superconducting Fermi liquid (FL) state. It is seen that the ambient-pressure superconductivity of the sample develops from a pure SM normal state, featured by $α=1$ (Fig.3b), and the sample holds the highest $T_c$ value. Upon increasing pressure till the critical pressure $P_c$, $T_c$ decreases monotonously, with $α$ varying between 1 and 2 (here, we take the average value of the pressures that destroy the superconducting state, obtained from different experimental runs, as $P_c$, see Fig.3). At $P_c$ (~12.5 GPa), the superconductivity is completely suppressed (Fig.3a) and the power exponent $α$ approaches 2 (Fig.3b), indicating that the ground state of the sample moves into a non-superconducting FL state. These results demonstrate that applying pressure renders the 1048 superconductor to undergo a quantum phase transition at ~12.5 GPa. Below the quantum critical point (QCP), the superconductor has a SC ground state with a SM normal state or NFL behavior, while above the QCP the sample displays a non-superconducting FL state. The pressure-induced quantum phase transition observed in the 1048 superconductor is highly reminiscent of what has been seen in the over-doped cuprate superconductors[2,3,5,6,26] and FeSe superconductor[32], which show the same evolution upon increasing doping level and gating voltage.

Furthermore, we track the change of the coefficient $A^{\square}$ as a function of pressure before reaching the QCP. As shown in Fig.3c, the coefficient $A^{\square}$ versus pressure displays a similar trend with $T_c$ versus pressure. As $A^{\square}$ holds the maximum, the 1048 superconductor displays the highest $T_c$ value. Once $A^{\square}$ is decreased by pressure, $T_c$ exhibits a decrease correspondingly. More significantly, we find that the values of the coefficient $A^{\square}$ and $T_c$ reach zero together at the QCP. The observation of the synchronized decrease of $A^{\square}$ and $T_c$ with applied pressure leads us to propose that the coefficient $A^{\square}$ is a key factor for determining the $T_c$ value of the high-$T_c$ superconductors.

To establish the correlation between the SM normal state and $T_c$ for the 1048 superconductor and compare it with doped cuprate and other superconductors, we plot the $T_c$ dependence of the coefficient $A^{\square}$ in Fig.4a and find a positive correlation of $A^{\square}$ with $T_c$. Scaling analysis on the two quantities finds that $T_c$ versus $A^{\square}$ basically obeys the relation of $T_c \sim (A^{\square})^{0.5}$ (see the inset of Fig.4a). To the best of our knowledge, this is the first quantitative description on the correlation between $T_c$ and $A^{\square}$ for the iron-pnictide superconductors through the pressure tuning method. For the sake of investigating the possible universality of the concurrent breakdown of the SM normal state and superconducting state, we have also performed hydrostatic pressure studies on $Sr_{0.74}Na_{0.26}Fe_2As_2$ superconductor, whose normal resistance versus temperature displays a FL state at ambient pressure, but turns into a SM state at 5.9 GPa (see SI). The results obtained also show the concurrent breakdowns of $A^{\square}$ and $T_c$ at the same pressure point (see SI). Further, we compare our results with the experimental data obtained from chemically-doped cuprate[6,8,26,33] and iron-based superconductors[34-35], and find the similar relationship between $T_c$ and $A^{\square}$ (Fig.4b), demonstrating that the stability of the superconducting phase has an intimate connection with the contribution from the electron state of determining the $T$-linear resistivity.

It has been proposed that antiferromagnetic (AFM) spin fluctuations are responsible for both the superconducting and SM normal states in the high-$T_c$ superconductors[33,36]. If the AFM fluctuations are the fundamental principle, then a theory to explain the detailed physics related to the behavior of the SM state, as revealed in this study, is also needed.

In conclusion, our results reveal that the application of pressure induces a quantum phase transition (QPT) from a SC ground state with a SM normal state to a non-superconducting FL state in the $Ca_{10}(Pt_4As_8)((Fe_{0.97}Pt_{0.03})_2As_2)_5$ superconductor. The observed pressure-tuned coevolution of the strange metal (SM) and superconducting (SC) states as well as the concurrent breakdown of the SM and SC states at the quantum critical pressure (~12.5 GPa) demonstrates how the coefficient $A^\square$, the slope of the linear-in-temperature resistivity normalized by the average distance between FeAs layers, determines the stability of the superconductivity and the value of $T_c$. Below the QCP, the coefficient $A^\square$ is continuously suppressed from a finite value and the power component $\alpha$ varies between 1 and 2. Above the QCP, the SC state is fully suppressed, while $A^\square = 0$ and $\alpha = 2$ simultaneously, a hallmark of a pure Fermi liquid (FL) state. We have also observed the same phenomenon in pressurized $Sr_{0.74}Na_{0.26}Fe_2As_2$ superconductor. The scaling analysis for the obtained $T_c$ and coefficient $A^\square$ in the compressed 1048 superconductor find that it basically obey the relation of $T_c \sim (A^\square)^{0.5}$, exhibiting the similarity with what is seen in other chemically-doped unconventional superconductors and demonstrating for the first time that the high-$T_c$ superconductors with a SM normal state obey the same relation in $T_c(A^\square)$, regardless of the type of the tuning method (doping or pressurizing), the crystal structure, the bulk or film superconductors and the nature of dopant (electrons or holes). The analysis on the generic relation between the $T_c$ and the SM state for the cuprate and iron-based high-$T_c$ superconductors reveals that the same physics governs the emergence and the stability of the high-$T_c$ superconductivity.

**References**

1. Zaanen, J. Planckian dissipation, minimal viscosity and the transport in cuprate strange metals. *SciPost Phys.* **6**, 061 (2019).
2. Greene, R.L. et al. The Strange Metal State of the Electron-Doped Cuprates. *Annu. Rev. Condens. Matter Phys.* **11**, 213-229 (2020).
3. Legros, A. et al. Universal T-linear resistivity and Planckian dissipation in overdoped cuprates. *Nat. Phys.* **15**, 142-147 (2019).
4. Jin, K. et al. Link between spin fluctuations and electron pairing in copper oxide superconductors. *Nature* **476**, 73-75 (2011).
5. Ayres, J. et al. Incoherent transport across the strange-metal regime of overdoped cuprates. *Nature* **595**, 661-666 (2021).
6. Bozovic, I. et al. Dependence of the critical temperature in overdoped copper oxides on superfluid density. *Nature* **536**, 309-311 (2016).
7. Martin, S. et al. Normal-state transport properties of $Bi_{2+x}Sr_{2-y}CuO_{6+\delta}$ crystals. *Phys. Rev. B Condens. Matter* **41**, 846-849 (1990).
8. Doiron-Leyraud, N. et al. Correlation between linear resistivity and Tc in the Bechgaard salts and the pnictide superconductor $Ba(Fe_{1-x}Co_x)_2As_2$. *Phys. Rev. B* **80**, 214531 (2009).
9. Boebinger Gregory, S. An Abnormal Normal State. *Science* **323**, 590-591 (2009).
10. Coleman, P. et al. Quantum criticality. *Nature* **433**, 226-229 (2005).
11. Park, T. et al. Isotropic quantum scattering and unconventional superconductivity. *Nature* **456**, 366-368 (2008).
12. Knebel, G. et al. The Quantum Critical Point in $CeRhIn_5$: A Resistivity Study. *J. Phys. Soc. Jpn.* **77**, 114704-114704 (2008).

13. Nguyen, D.H. et al. Superconductivity in an extreme strange metal. *Nat. Commun.* **12**, 4341 (2021).

14. Li, D. et al. Superconductivity in an infinite-layer nickelate. *Nature* **572**, 624-627 (2019).

15. Cao, Y. et al. Strange Metal in Magic-Angle Graphene with near Planckian Dissipation. *Phys. Rev. Let.t* **124**, 076801 (2020).

16. Ni, N. et al. High Tc electron doped $Ca_{10}(Pt_3As_8)(Fe_2As_2)_5$ and $Ca_{10}(Pt_4As_8)(Fe_2As_2)_5$ superconductors with skutterudite intermediary layers. *Proc. Natl. Acad. Sci. U.S.A.* **108**, E1019-E1026 (2011).

17. Mao, H.-K. et al. Solids, liquids, and gases under high pressure. *Rev. Mod. Phys.* **90**, 015007 (2018).

18. Gao, L. et al. Superconductivity up to 164 K in $HgBa_2Ca_{m-1}Cu_mO_{2m+2+\delta}$ (m=1, 2, and 3) under quasihydrostatic pressures. *Phys. Rev. B* **50**, 4260-4263 (1994).

19. Takahashi, H. et al. Superconductivity at 43 K in an iron-based layered compound $LaO_{1-x}F_xFeAs$. *Nature* **453**, 376-378 (2008).

20. Torikachvili, M.S. et al. Pressure induced superconductivity in $CaFe_2As_2$. *Phys. Rev. Lett* **101**, 057006 (2008).

21. Deng, L. et al. Higher superconducting transition temperature by breaking the universal pressure relation. *Proc. Natl. Acad. Sci. U.S.A.* **116**, 2004-2008 (2019).

22. Shimizu, K. et al. Superconductivity in compressed lithium at 20 K. *Nature* **419**, 597-599 (2002).

23. Zhou, Y. et al. Quantum phase transition from superconducting to insulating-like state in a pressurized cuprate superconductor. *Nat. Phys.* **18**, 406-410 (2022).

24. Sun, L. et al. Re-emerging superconductivity at 48 kelvin in iron chalcogenides. *Nature* **483**, 67-69 (2012).

25. Guo, J. et al. Pressure-driven quantum criticality in iron-selenide superconductors. *Phys. Rev. Lett.* **108**, 197001 (2012).

26. Yuan, J. et al. Scaling of the strange-metal scattering in unconventional superconductors. *Nature* **602**, 431-436 (2022).

27. Proust, C. et al. The Remarkable Underlying Ground States of Cuprate

Superconductors. *Annu. Rev. Condens. Matter Phys.* **10**, 409-429 (2019).

28. Li, Q. et al. Two-Dimensional Superconducting Fluctuations in Stripe-Ordered La$_{1.875}$Ba$_{0.125}$CuO$_4$. *Phys. Rev. Lett.* **99**, 067001 (2007).

29. Meier, W.R. et al. Optimization of the crystal growth of the superconductor CaKFe$_4$As$_4$ from solution in the FeAs-CaFe$_2$As$_2$-KFe$_2$As$_2$ system. *Phys. Rev. Mater.* **1**, 013401 (2017).

30. Chu, C.W. et al. Hole-doped cuprate high temperature superconductors. *Physica C: Superconductivity and its Applications* **514**, 290-313 (2015).

31. Wu, M.K. et al. Superconductivity at 93 K in a new mixed-phase Y-Ba-Cu-O compound system at ambient pressure. *Phys. Rev. Lett.* **58**, 908-910 (1987).

32. Jiang, X. et al. Interplay between superconductivity and the strange-metal state in FeSe. *Nature Physics* (2023, https://doi.org/10.1038/s41567-022-01894-4)

33. Taillefer, L. Scattering and Pairing in Cuprate Superconductors. *Annu. Rev. Condens. Matter Phys.* **1**, 51-70 (2010).

34. Chu, J.-H. et al. Determination of the phase diagram of the electron-doped superconductor Ba(Fe$_{1-x}$Co$_x$)$_2$As$_2$. *Phys. Rev. B* **79**, 014506 (2009).

35. Fang, L. et al. Roles of multiband effects and electron-hole asymmetry in the superconductivity and normal-state properties of Ba(Fe$_{1-x}$Co$_x$)$_2$As$_2$. *Phys. Rev. B* **80**, 140508(R) (2009).

36. Sedeki, A. et al. Extended quantum criticality of low-dimensional superconductors near a spin-density-wave instability. *Phys. Rev. B* **85,** 165129 (2012).

**Acknowledgements**

This work in China was supported by the NSF of China (Grant Numbers Grants No. U2032214, 12104487, 12122414 and 12004419), the National Key Research and Development Program of China (Grant No. 2021YFA1401800), and the Strategic Priority Research Program (B) of the Chinese Academy of Sciences (Grant No. XDB25000000). We thank the support from the Users with Excellence Program of Hefei Science Center CAS (2020HSC-UE015). J. G. and S.C. are grateful for supports from the Youth Innovation Promotion Association of the CAS (2019008) and the China Postdoctoral Science Foundation (E0BK111). The work at Princeton was supported by the US Department of Energy, Division of Basic Energy Sciences, grant DE-FG02-98ER-45706. Work at UCLA was supported by the U.S. Department of Energy (DOE), Office of Science, Office of Basic Energy Sciences under Award Number DE-SC0021117.

**Author information**

The authors declare no competing financial interest. Correspondence and requests for materials should be addressed to L.S.(llsun@iphy.ac.cn).

**Data availability**

The data that support the findings of this study are available from the corresponding author upon reasonable request.

**Author contributions**

L.S., T.X. and Q.W. designed the study and supervised the project. N.N. and J.R.C. grew the $Ca_{10}(Pt_4As_8)(Fe_2As_2)_5$ single crystals. R.Y and X.G.Q. grew the $Sr_{0.74}Na_{0.26}Fe_2As_2$ single crystals. S.C. , J.Y.Z., J.G., P.Y.W., J.Y.H., S.J.L, Y.Z.Z. and L.S. performed the high pressure resistance measurements. L.S., T.X., Q.W. S.C. and R.J.C wrote the manuscript in consultation with all authors.

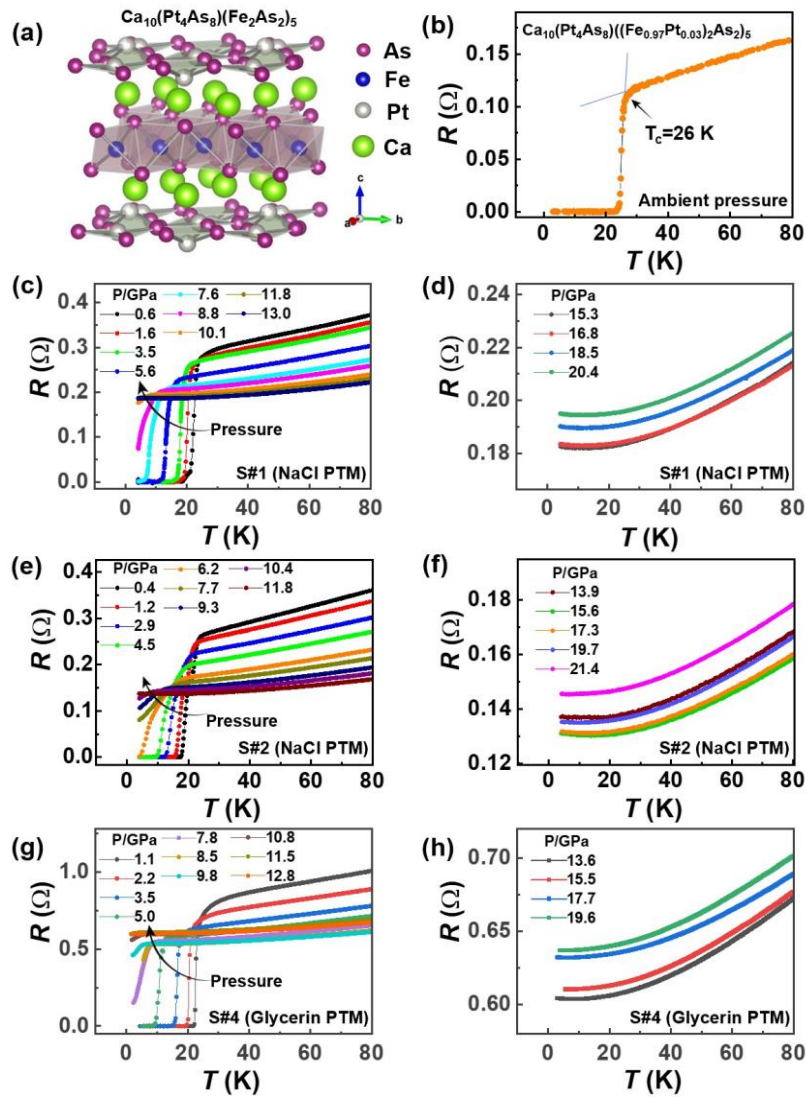

**Figure 1 Structure information and transport properties for the 1048 superconductor.** (a) The crystal structure of the 1048 superconductor. (b) Ambient-pressure resistance as a function of temperature, displaying a superconducting transition with an extradentary sharp drop at 26 K. (c) - (f) Temperature dependence of resistance for the S#1 and S#2 surrounded by the pressure transmitting medium (PTM) of NaCl in the measurements. (g) and (h) Resistance versus temperature for the S#4 surrounded by the liquid PTM of glycerin in the measurements.

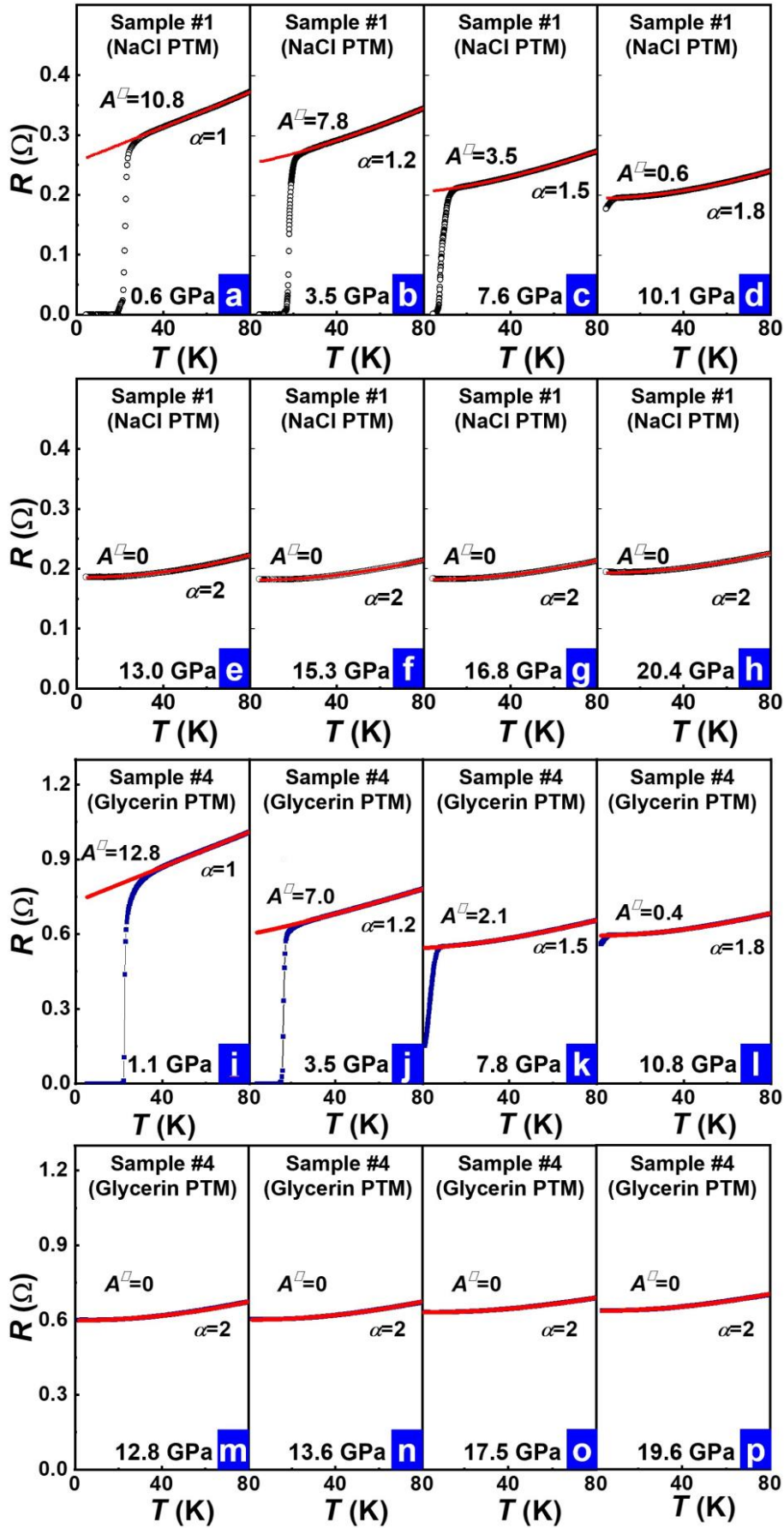

**Figure 2 High-pressure resistance as a function of temperature at different pressures for the $Ca_{10}(Pt_4As_8)((Fe_{0.97}Pt_{0.03})_2As_2)_5$ superconductors and corresponding fits by the forms of $R = R_0' + A'T + B'T^2$.** (a) $A^{\square}$ ($A^{\square}=(w/l)(t/d)A'$, here $w$ and $t$ are the width and the thickness of the sample, $l$ is the distance between the electrodes, and $d$ is an average distance between FeAs layers, see SI) shows the maximum value (10.8) at 0.6 GPa for the S#1 surrounded by the pressure transmitting medium (PTM) of NaCl, where the normal state of sample is in a pure SM state ($\alpha=1$) and its $T_c$ displays the highest value. (b)-(d) $A^{\square}$ decreases with applied pressure in the pressure range of 3.5 GPa - 10.1 GPa for the S#1, in the pressure range of which $T_c$ is suppressed gradually. (e-h) $A^{\square}$ approaches zero and $\alpha=2$ for the S#1 in the pressure range of 13.0 GPa - 20.4 GPa, in which the superconductivity is absent. (i) $A^{\square}=12.8$ at 1.1 GPa for the S#4 surrounded by the PTM of glycerin, where the normal state of sample is also in a pure SM state ($\alpha=1$) and its $T_c$ displays the highest value. (j-l) $A^{\square}$ declines upon increasing pressure for the S#4 in the pressure range of 3.5 GPa - 10.8GPa. (m-p) $A^{\square}=0$ and $\alpha=2$ for the S#4 in the pressure range of 12.8 GPa - 19.6 GPa, in which the sample is no longer superconducting.

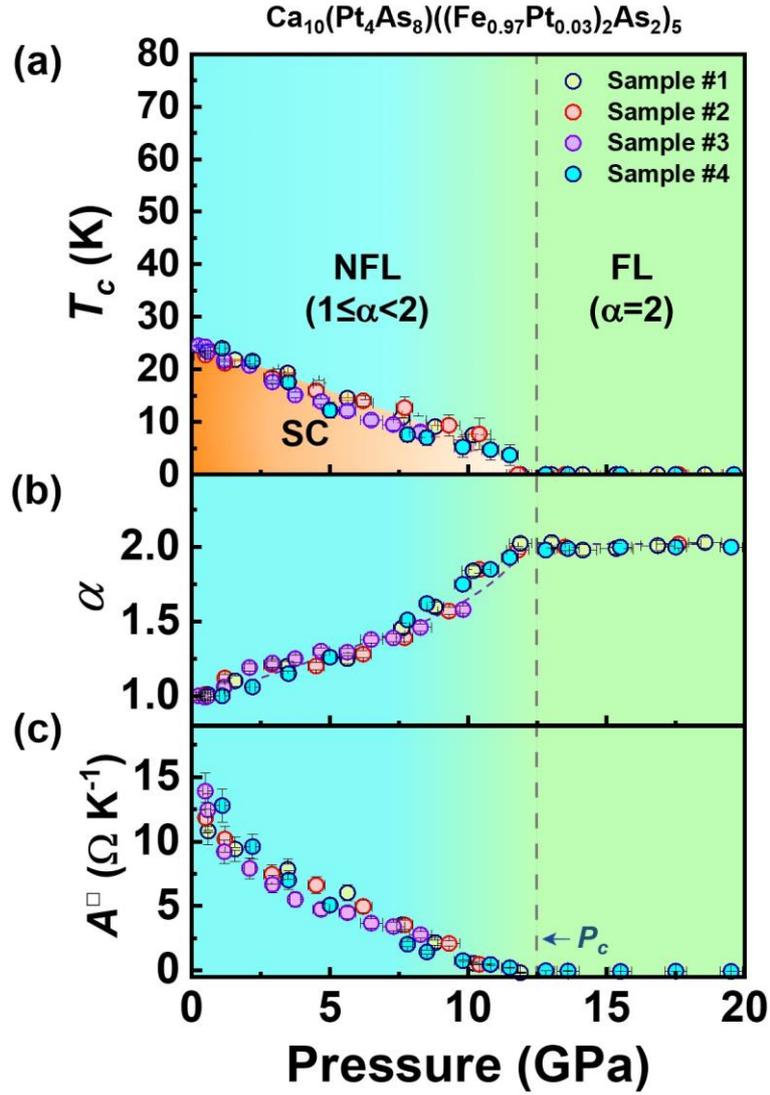

**Figure 3 Summary of the superconductivity, power exponent ($\alpha$) and the coefficient ($A^{\square}$) as a function of pressure for the $Ca_{10}(Pt_4As_8)((Fe_{0.97}Pt_{0.03})_2As_2)_5$ superconductor.** (a) Pressure-$T_c$ phase diagram. NFL and FL represent the non-Fermi liquid behavior and Fermi liquid state, respectively. SC stands for superconducting state. The $T_c$ value in the phase diagram is determined by the onset transition temperature. (b) The plot of $\alpha$ in the form of $R(T) = R_0 + AT^{\alpha}$ versus pressure. (c) Pressure dependence of the coefficient $A^{\square}$.

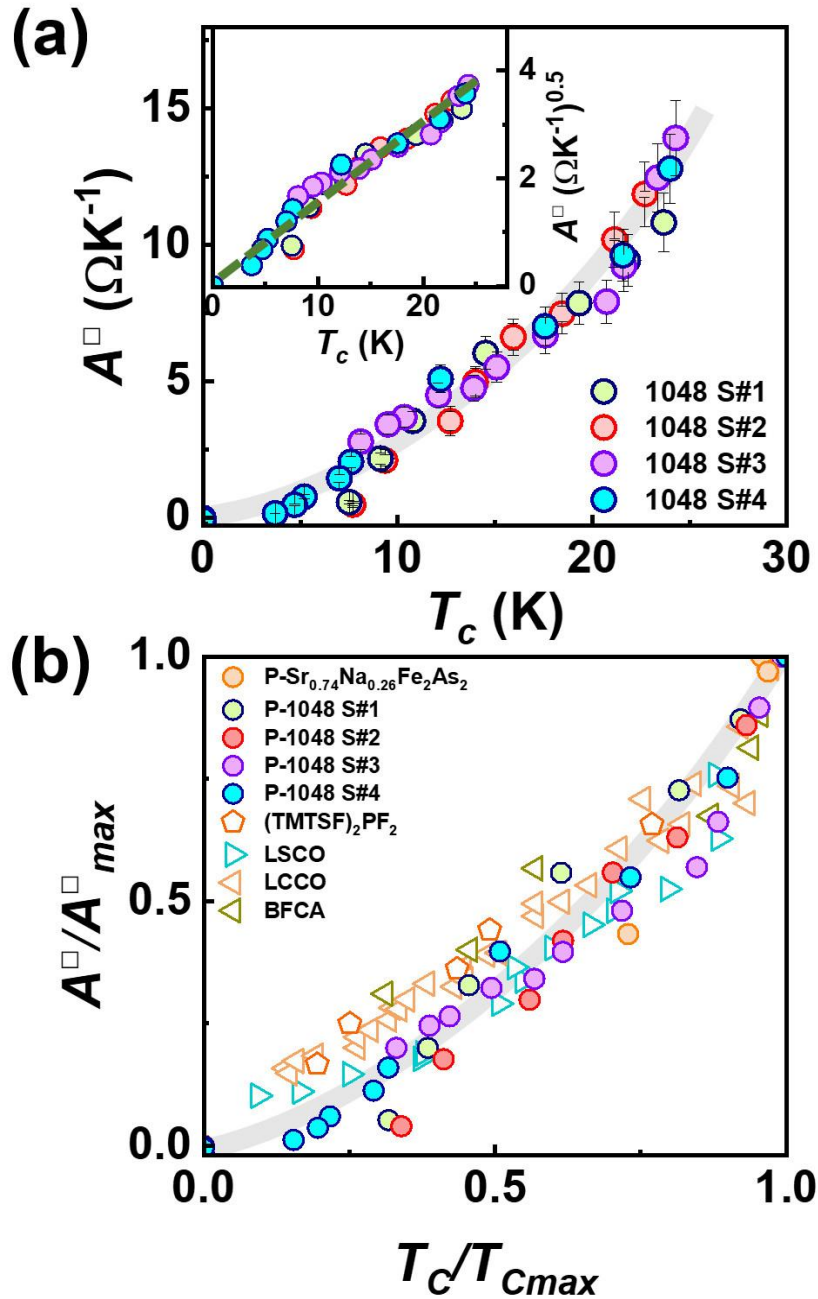

**Figure 4 Superconducting transition temperature versus slope of linear-in-temperature resistivity for the pressurized $Ca_{10}(Pt_4As_8)((Fe_{0.97}Pt_{0.03})_2As_2)_5$ superconductor.** (a) The coefficient $A^{\square}$ as a function of superconducting transition temperature ($T_c$). The inset displays $T_c$ dependence of $(A^{\square})^{0.5}$. (b) The summary of the correlation between $A^{\square}/A^{\square max}$ and $T_c/T_c^{max}$ for the pressurized $Ca_{10}(Pt_4As_8)((Fe_{0.97}Pt_{0.03})_2As_2)_5$ and $Sr_{0.74}Na_{0.26}Fe_2As_2$ superconductors, as well as

other differently-doped superconductors. The LCCO, LSCO and BFCA represent $La_{2-x}Ce_xCuO_4$ superconductor [Ref.26], $La_{2-x}Sr_xCuO_4$, superconductor [Ref. 6 and Ref.26] and $Ba(Fe_{1-x}Co_x)_2As_2$ superconductor [Ref.26 and Ref.34-35], respectively. The data about $(TMTSF)_2PF_2$ are taken from Ref. 8.